\documentclass[conference]{IEEEtran}
\IEEEoverridecommandlockouts
\usepackage{cite}
\usepackage{amsmath,amssymb,amsfonts}
\usepackage{algorithmic}
\usepackage{listings}
\usepackage{subfigure}
\usepackage{graphicx}
\usepackage{url}
\usepackage{textcomp}
\usepackage{xcolor}
\usepackage{geometry}
\definecolor{codegreen}{rgb}{0,0.6,0}
\definecolor{codegray}{rgb}{0.5,0.5,0.5}
\definecolor{codepurple}{rgb}{0.58,0,0.82}
\definecolor{backcolour}{rgb}{0.95,0.95,0.92}

\lstdefinestyle{mystyle}{
    backgroundcolor=\color{white},   
    commentstyle=\color{blue},
    keywordstyle=\color{red},
    numberstyle=\tiny\color{codegray},
    stringstyle=\color{codepurple},
    basicstyle=\ttfamily\footnotesize,
    breakatwhitespace=false,         
    breaklines=true,                 
    captionpos=b,                    
    keepspaces=true,                 
    numbers=left,                    
    numbersep=0pt,    
    showspaces=false,                
    showstringspaces=false,
    showtabs=false,  
    tabsize=1
}

\def\BibTeX{{\rm B\kern-.05em{\sc i\kern-.025em b}\kern-.08em
    T\kern-.1667em\lower.7ex\hbox{E}\kern-.125emX}}

\begin{document}

\title{\huge \textbf{\textsc{Holmes}: An Efficient and Lightweight Semantic Based Anomalous Email Detector}}


\author{\IEEEauthorblockN{Peilun Wu\IEEEauthorrefmark{1} and Hui Guo\IEEEauthorrefmark{4}}
\IEEEauthorblockA{
Data Security \& Compliance, CDO Data \& Cloud, PwC CN.\IEEEauthorrefmark{1}\\
School of Computer Science and Engineering, University of New South Wales (UNSW)\IEEEauthorrefmark{1}\IEEEauthorrefmark{4}}
Email: \IEEEauthorrefmark{1}z5100023@zmail.unsw.edu.au,
\IEEEauthorrefmark{4}h.guo@unsw.edu.au}

\maketitle

\begin{abstract}
Email threat is a serious issue for enterprise security. The threat can be in various malicious forms, such as phishing, fraud, blackmail and malvertisement.
The traditional anti-spam gateway often maintains a greylist to filter out unexpected emails based on suspicious vocabularies present in the email's subject and contents. 
However, this type of signature-based approach cannot effectively discover novel and unknown suspicious emails that utilize various evolving malicious payloads.
To address the problem, in this paper, we present \textsc{Holmes}, an efficient and lightweight semantic based engine for anomalous email detection.
\textsc{Holmes} can convert each email event log into a sentence through word embedding and then identify abnormalities that deviate from a historical baseline based on those translated sentences. 
We have evaluated the performance of \textsc{Holmes} in a real-world enterprise environment, where around 5,000 emails are sent/received each day.
In our experiments, \textsc{Holmes} shows a high capability to detect email threats, especially those that cannot be handled by the enterprise anti-spam gateway.
It is also demonstrated through our experiment that \textsc{Holmes} can discover more concealed malicious emails that are immune from several commercial detection tools.
\end{abstract}


\section{Introduction}
\label{sec:introduction}

Though the instant messaging software, such as Facebook and WeChat, has gained increasing popularity, the email service is still indispensable for enterprises.
Since the email service is a public-facing application, it can be targeted by the hacker as an easy entrance to the internal network.     
Based on our observations, fraud, malvertisement and spread-phishing are the main email threats frequently received by enterprise users. 
These emails use deceptive subjects to pretend and hide themselves.
Usually, malware infected attachments or malicious URLs are embedded in the email body to spoof recipients for further action.
Once the attachment is downloaded or a link is clicked, the recipients's system is compromised or the confidential information is leaked \cite{foster2015security}.

To alleviate the problem, an enterprise often deploys some anti-spam gateways to filter out unexpected emails.
However, the associated techniques for spam detection, such as greylist and subject analysis, cannot effectively discover novel and unknown email threats that are elaborately constructed by utilizing various current hot topics, such as COVID-19, US election.
These unknown threats can easily bypass the anti-spam gateway and successfully permeate the target system, leading to a series of damaging consequences, such as administrator account theft, database attack and financial blackmail.

In this paper, we introduce a novel artificial intelligence based anomalous email detector, \textsc{Holmes}, that can effectively tackle the challenges mentioned above.
\textsc{Holmes} combines word embedding with novelty detection to discover anomalous behaviours from a high volume of mirrored SMTP traffic in a large-scale enterprise environment. 
To improve the result interpret-ability, we trace the real source IP addresses of suspicious emails in line with their geographical positions and further visualize the correlated relations in a directed-force graph.
Our contributions are summarized as follows:
\begin{itemize}
    \item We propose an efficient and lightweight semantic based anomalous email detector, \textsc{Holmes}.
    Different from other detectors that usually require to examine email bodies, \textsc{Holmes} can discover anomalies simply based on email headers, which significantly reduces the cost of resource consumption and avoids accessing email bodies (a sensitive security issue).
    \item We exploit graph visualization to reveal the correlated relations of detected suspicious emails
    and demonstrate that the attacker portrait (based on their geographical positions) is in line with the cyber threat intelligence provided.  
    \item We evaluate \textsc{Holmes} with a commercial anti-spam gateway deployed in a real-world enterprise environment.
    \textsc{Holmes} not only can accurately detect those email threats that have been blocked by the anti-spam gateway, but also can discover a large number of email threats that have successfully escaped from the gateway.  
    We also compare \textsc{Holmes} with several commercial email detectors offered by different security vendors in VirusTotal\cite{VirusTotal}, which shows that \textsc{Holmes} outperforms those detectors with a very high detection rate on the use of threat hunting in the wild. 
\end{itemize}

The remainder of the paper is structured as follows. 
We begin with a brief discussion of some related work on email detection in Section~\ref{sec:background}. 
We then in Section~\ref{sec:Holmes} introduce the proposed semantic based anomalous email detector, \textsc{Holmes}. 
In Section~\ref{sec:evaluation}, we present our evaluation results of \textsc{Holmes} and several commercial security 
products; a demonstration of how visualization can be used to reconstruct the attack stories is also given in this section. The enhancements on \textsc{Holmes} for the real world implementation is given in Section~\ref{sec:implementation}.
The paper is concluded in Section~\ref{sec:conclusion}. 
As an add-on section, we append some extra discussions at the end of this paper.

\section{Preliminary Knowledge}
\label{sec:background}
Anomalous emails can be classified into external threats and internal threats in accordance with MITRE ATT\&CK Matrix~\cite{MITRE}.
External threats are the emails sent from external sources, whereas the internal threats are the emails sent from 
legitimate users within an organization but whose email accounts have been stolen and used for the lateral movement attack. 
Most of previous research mainly focuses on one specific threat type, such as URL-based lateral phishing~\cite{ho2019detecting} or phishing web pages from search engine in a large-scale cyberspace~\cite{whittaker2010large}.  
There are still many open questions and unsolved challenges that need to be addressed holistically.
Some issues and the existing solutions are presented below.\\

\noindent\textbf{No Built-In Authentication in SMTP.}
The lack of a native authentication mechanism inside the SMTP service presents a security loophole to attackers. Attackers
can easily forge the email header by pretending to be someone the recipient knows or from a business the recipient has a relationship with, so as to spoof recipients and avoid spam block lists~\cite{Barracuda}.
To address the problem, several frameworks, such as SPF\cite{wong2006sender} (Sender Policy Framework), DKIM\cite{allman2007domainkeys} (Domain Key Identified Mail) and DMARC\cite{kucherawy2015domain} (Domain-Based Message Authentication, Reporting, and Conformance),  have been developed to incorporate authentication into the email system. However, these designs are still not very effective in terms of implementation. When integrating authentication into the mail system with a typical component-based software design, there are inconsistency issues between the software components offered by different parties~\cite{chen2020composition}, such as the incompatibility of mail forwarding servers, which allows numerous email threats escape detection. \\


\noindent\textbf{Lack of Sensitivity to Unknown Variations.}
The unreliability of SMTP leaves email threats to have evolved into many variations, which are difficult to be discovered by the traditional security products.
We have evaluated several malicious email detection modules within our internal security products that use pattern matching of attack signatures for anomaly detection.
None of them can discover the crafted phishing emails that utilize business-related content to pretend themselves look normal for evasion. 
We also have used the crafted phishing samples collected from our real-world hunting to evaluate the detection rate of 60+ typical detection engines in VirusTotal (Enterprise Service).
Nevertheless, the evaluation result also shows their low sensitivity to unknown threats -- in fact, all testing samples can successfully escape from the detection of those engines.
This kind of low ability of detecting unknown attack variations has motivated the security community to turn to AI-based methods for anomaly detection.\\

\noindent\textbf{High False Positive Rate.}
The research on anomaly detection for cyber threat hunting has been around for decades.
The main concern on applying machine learning for anomaly detection is the significant false positive rate (FPR).
Even though new designs are continuously proposed aiming for improvement\cite{wu2020densely,wu2020pelican,wu2019lunet}, they were rarely evaluated in a real-world working environment, let alone put into use in commercial systems.\\


\noindent\textbf{High Cost and Performance Bottleneck.}
The imbalance between the cost of data collection and the performance of algorithmic consumption is a significant challenge for most of the AI-based detectors. 
Though the complexity of AI computing algorithms  has been constantly improved, most AI modules still require large computing and storage resources. which makes the existing attack detectors not easy to use and very slow to response attacks.
Furthermore, the detectors that use supervised machine learning require 
the labeled input data records
and often need to be retrained once their performance begins to degrade, which also makes the machine learning ineffective for detection automation.
\\

\noindent\textbf{Lack of Provenance Analysis.}
So far few detectors have considered to integrate the provenance analysis within the detection mechanism.
We believe provenance analysis is an important and enabling component in malicious email detection.
Provenance analysis\cite{hassan2020tactical} can reveal the attack story and the detail of attacker portrait behind the email, such as (1) where the email is from, (2) who the real sender is, (3) how the malicious shellcode executes, (4) what the potential correlations between malicious events are.
The above information is important for the security team to analyze the attack techniques, tactics and procedures (TTPs) and further assist the security experts to identify individual attackers or organizations.

\begin{figure}[t]
    \centering
    \includegraphics[width=0.98\linewidth]{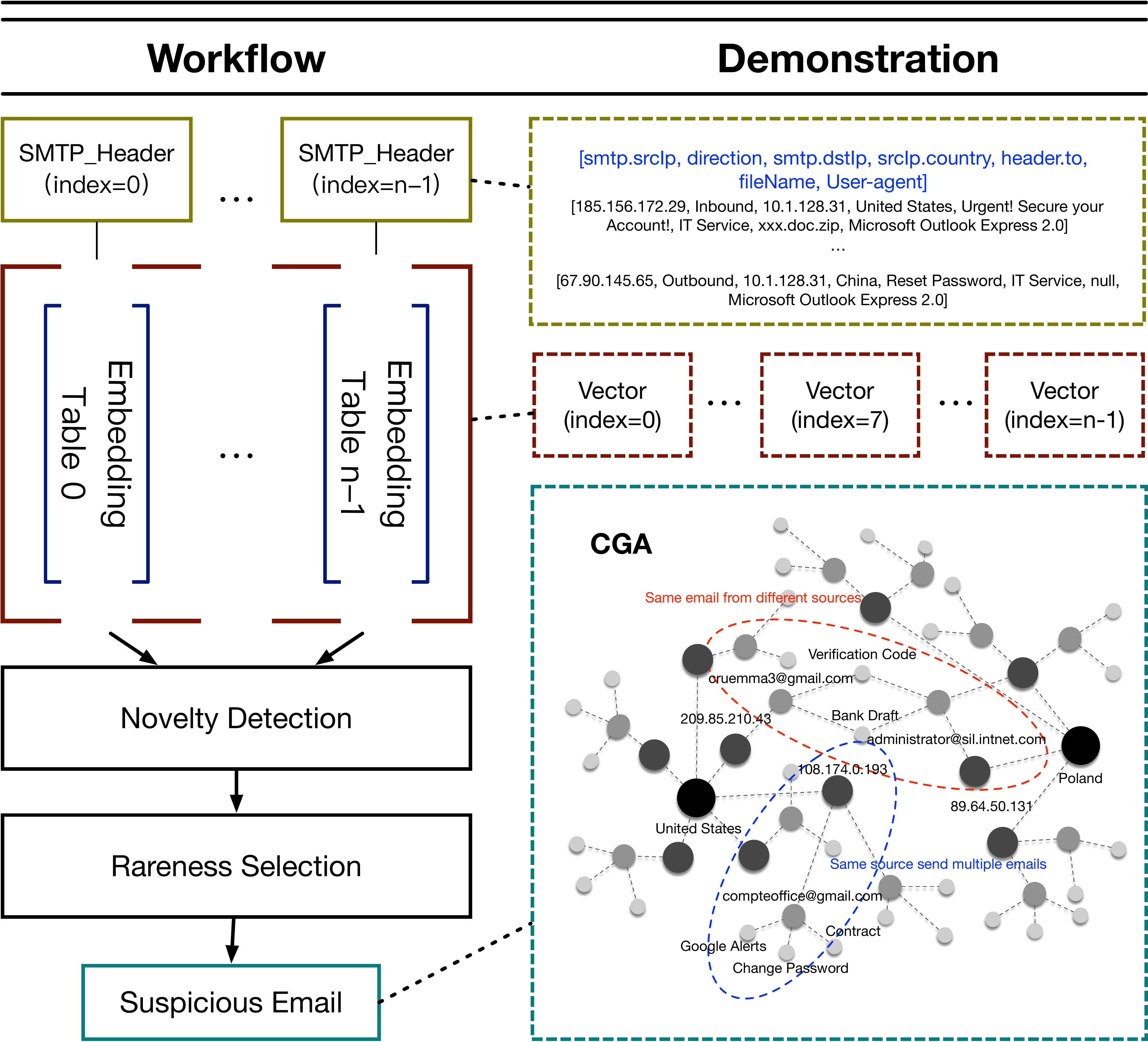}
    \caption{\textsc{Holmes}' workflow and demonstration}
    \label{fig:Holmes}
\end{figure}

\section{\textsc{Holmes} - Anomalous Email Detector}
\label{sec:Holmes}
To address the above challenges, we introduce an efficient
and lightweight semantic oriented anomalous email
detector, \textsc{Holmes}, that can detect an email attack by
analyzing the sender-recipient relation, which is available in
the email header of SMTP.

\textsc{Holmes} is a threat hunting tool for the incident response and investigation. It works on mirrored network traffic to inspect and report anomalies in the bypass but do not block them directly.
\textsc{Holmes} is used to assist the incident response team to discover more concealed threats that can escape from the detection of traditional anti-spam gateway and threaten the network security.

\textsc{Holmes} is a self-adaptive learning machine. It can learn historical SMTP traffic from last 24 hours and detect anomalies that deviate from the baseline of historical behaviour in the next 24 hours.
Fig~\ref{fig:Holmes} shows the overall \textsc{Holmes}  workflow along with the demonstrations as how data/information is evolved in the system.
\textsc{Holmes} consists of four main functions: 
1) Word Embedding, 2) Novelty Detection, 3) Rareness Selection, and 4) Correlation Graph Analysis.
For each email, \textsc{Holmes} takes its  header and converts it into a numeric presentation through word embedding. The numeric data records are then input to the machine learning unit (Novelty Detection). The unit generates a list of novel emails, which are, in turn, processed by the rareness selection procedure to narrow down the detection targets. The detected results are finally presented in a human readable format and the correlations of the related email attacks are also pictured with a graph. 

\textsc{Holmes} is also a super lightweight and high-efficient detector, which was originally written in Python with only 52 lines of code. 
Based on the run-time analysis, \textsc{Holmes} can complete the entire detection in less than 73 seconds with 127 MB memory consumption on around 700 MB datasets.
In this section, we would open-source the original code and detail the implementation of the main detection functions in a hope to assist researchers to evaluate and reuse \textsc{Holmes} in their future research.  

\subsection{Word Embedding}
Since the email textual header information cannot be directly used for machine learning, \textbf{how to effectively represent textual data to the machine understandable} is important.
Most algorithm engineers use OneHotEncoder\cite{haq2018categorical} or OrdinalEncoder\cite{cerda2018similarity}, or Bag-of-Words (BOW)\cite{zhang2010understanding},
which can be simply implemented by the open-sourced library Scikit-Learn.
However, those methods are not able to effectively maintain the data semantic correlations in either temporal or in spatial dimension.

To address the problem, we use paragraph vector (Doc2Vec)\cite{le2014distributed} for the conversion, as detailed by the Python code in Listing 1. 

\begin{lstlisting}[language=Python, caption=Doc2Vec]
    def Doc2Vec(self,feature):
        '''
        :description: convert textual SMTP header to vectors
        :param feature: SMTP features
        :return: word vectors
        '''
        # build a vocabulary for each feature
        documents = [TaggedDocument(doc, [i]) for i, doc in enumerate(feature)]
        # set vector size
        model = Doc2Vec(vector_size=40, min_count=2, epochs=40)
        # build model
        model.build_vocab(documents) 
        # train the model
        model.train(documents, total_examples=model.corpus_count, epochs=model.epochs)
        # return vectors
        return model.docvecs.vectors_docs
\end{lstlisting}

Compared to other conversion methods, Doc2Vec is able to better keep the semantics of the words or more formally the distances between the words, which can be of variable-length ranging from sentences to documents.
Doc2Vec is a semi-supervised learning algorithm. Its input is unlabeled but what will be learned is specified/supervised. 
In our code, the inputs are email headers and what to be learned are the features in the header, as highlighted in blue in the top-right block in Fig.~\ref{fig:Holmes}. Besides some basic attributes such as subject, header.from or user-agent, which are often forged by hackers, we design two additional features that can also be used to help identify anomalies: the direction of email (direction) and the country of source IP address (srcIp.country).
The code in Listing 1 converts each email event to a feature vector, as illustrated in the right side middle boxes in Fig.~\ref{fig:Holmes}. 
The feature vectors are then used for novelty detection.

\subsection{Novelty Detection}
\textbf{Anomalous emails are usually unknown and novel.} Their behaviors often deviate from the trace of historical normal activities.
We use Local Outlier Factor (LOF)\cite{kriegel2009loop} 
to discover those emails, as given in Listing 2 and Listing 3. 

\begin{lstlisting}[language=Python, caption=Local Outlier Factor (LOF)]
    def local_outlier_factor(self,train_feature,test_feature):
        '''
        :description: estimate the decision score for each event
        :param train_feature: training data
        :param test_feature: testing data
        :return decision scores
        '''
        # initialize the LOF parameters (not sensitive)
        LOF = LocalOutlierFactor(n_neighbors=20, novelty=True, contamination=0.5)
        # fit training data
        LOF.fit(train_feature)
        # return a list of decision scores of the testing data
        return list(LOF.decisilst:listing 2on_function(test_feature))
\end{lstlisting}

The LOF algorithm shown in Listing 2 can learn the feature vectors of historical emails (i.e. the train\_feature dataset in the code) then provide the outlier score for newly seen emails (from the test\_feature dataset).

There are some compelling advantages of applying LOF for novelty detection: (1) It allows to train learning model on the data with contamination;
(2) It has a low computing complexity and can be used for online-learning, hence avoiding the performance degradation and the cost of retraining; (3) It is not sensitive to fine-tuning, which is beneficial to the effectiveness and stability of parametric learning.

\begin{lstlisting}[language=Python, caption=Novelty Analysis]
    def novelty_analysis(self,factor,test_feature):
        '''
        :description: filter novel events in accordance with the decision scores
        :param factor: decision scores of LOF
        :param test_feature: testing data
        :return novel/unseen samples
        '''
        # initialize threshold as zero
        threshold = 0
        # initialize an empty list to store the index of outliers 
        outliers = []
        # initialize an empty list to store novelties
        novelties = []
        # if negative values return the index of outliers
        for score in factor:
            if score < threshold:
                outliers.append(factor.index(score))
        # return novelties according to the index of outliers
        for index in outliers:
            novelties.append(test_feature[index])
        return novelties
\end{lstlisting}

The decision scores from the LOF code can be negative and positive. The negative values indicate the abnormalities and the positive values indicate the normal behaviours.
We regard any vector with a score smaller than a threshold is associated with an anomalous email, which is traced by the its index in the dataset, as shown in Listing 3. 



\subsection{Rareness Selection}

If we consider the relation of sender and recipient in emails, anomalous emails are often associated with 
a weak relation in that \textbf{a hacker usually does not send and reply to an email with the same recipient regularly}.
We therefore can further narrow down the malicious emails based on the sender-recipient relation -- namely, those abnormal emails with a weak sender-receiver relation will be selected as the final detection result, as described in Listing 4.

\begin{lstlisting}[language=Python, caption=Rareness Selection]
     def rareness_selection(self,focus_data):
        '''
        :description: filter rare events from novel emails 
        :param data: unseen samples
        :return rare emails
        '''
        # initialize an empty dict to store the relation
        relation = {}
        # initialize an empty list to store the rareness events
        rareness = []
        # initialize the features of a relation: 
        for each_data in focus_data:
            src_ip = each_data[0]
            mail_from = each_data[1]
            mail_to = each_data[2]
            direction = each_data[3]
            # define the features of a relation
            each_relation = src_ip + direction + mail_from + mail_to
            # count the relation
            relation.setdefault(each_relation, []).append(each_data)
        # filter the rare events that are smaller than a fine-tuned threshold from the relation dict
        threshold = 2
        for k, v in relation.items():
            if len(v) <= threshold:
                rareness.extend(v)

\end{lstlisting}

In our design, the relationship is measured based on the combination of a set of email features: source IP (src\_ip), direction, sender (mail\_from), and receiver (mail\_to). For a strong sender-receiver relation, there should be many emails of the same IP-direction-sender-receiver value. Therefore, we 
count emails for different IP-direction-sender-receiver values and select those that have a low count value (smaller than a threshold) as malicious emails, where the threshold is often fine-tuned based on different network environment by the operation team. 

\subsection{Correlation Graph Analysis (CGA)}\label{CGA}
Most of prior research overlooked a problem: \textbf{what is the relation within the anomalies?}
Lack of an effective solution significantly increases the load of security analysts, blurs the attacker portraits, and further makes the provenance analysis difficult.
To address the problem, we introduce a correlation graph analysis (CGA) module to improve the clarity of attacker portrait descriptions by correlating different anomalous events.

CGA is a directed-force graph and in our design, each node consists of the selected header features: country, srcIp, sender and subject.
The directed graph enforces the nodes that have dense connections come closer but separates the nodes if they do not or have sparse connections. 
The graph depicts the similarity of different anomalies (such as the same srcIp, same subject or same sender) and centralizes the cluster in line with their geographical locations, hence significantly improving the interpret-ability of provenance analysis.

The graph on the bottom right of Fig.~\ref{fig:Holmes} demonstrates the visualization result of CGA, where two clusters (one in red and one in blue) highlight the connected components that are centralized in accordance with the country of srcIp.  
The blue cluster shows that the same malicious email but sent from different sources, and the red cluster reveals the same source sends multiple different malicious emails.

The CGA module can be used to generate active IOCs (Indicator of Compromise) for the Cyber Threat Intelligence Platform, where we can match the similar or same malicious incidents occurred to other customers based on the IOCs.

\section{Evaluation}
\label{sec:evaluation}
\textsc{Holmes} has been deployed in an enterprise environment, where it can read mirrored SMTP records from the Elastic-Search (ES) server.
In this section, we first present some case studies on the malicious emails detected by \textsc{Holmes} and then show the comparison result \textsc{Holmes} with other popular commercial email detectors. 


\subsection{Case Studies}

According to our monthly email system data, \textsc{Holmes} can discover around 1,000 anomalous emails each day. Among them, about 23\% are truly malicious. And most of the malicious emails contain either phishing links or malware infected attachments. 
The rest are mainly spams and only a few are false positives.


Based on the detection results, 
we derive some malicious emails from our email server, which were not blocked by the anti-spam gateway but have been identified by our security analysts as the high risks, to reconstruct the attack stories.
Here, we would showcase some delicate crafted phishing emails and describe their malicious behaviour in detail.

\subsubsection*{\textbf{Case A: Fake DHL Delivery Message}}

\begin{figure}[t]
    \centering
    \includegraphics[width=.95\linewidth]{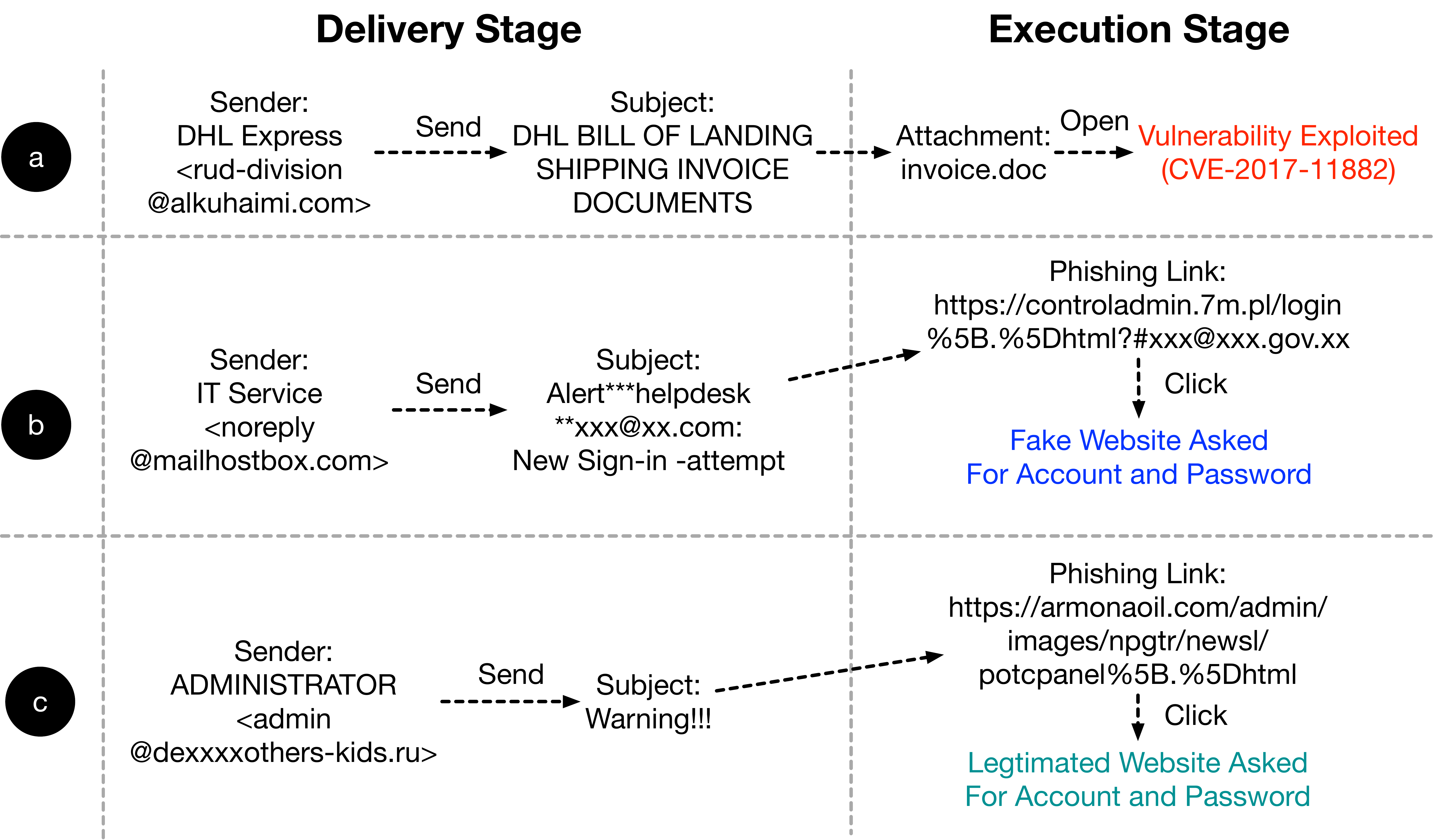}
    \caption{Case Studies}
    \label{fig:mail_samples}
\end{figure}

Fig.~\ref{fig:mail_samples}(a) shows the execution flow of a malicious email that pretends DHL service and plays following tricks: (1) The email uses a normally-seen subject that is associated with an invoice document; (2) The sender information has been modified as `DHL Express', which can be implemented by some hacking tools, such as swaks\cite{Swaks} or cobalt strike\cite{CobaltStrike}; (3) The email includes an attachment named \textit{invoice.doc}, which is, in fact, a malicious Trojan document that utilizes the CVE-2017-11882\cite{CVE-2017-11882} vulnerability; (4) The email contains a dedicated picture of DHL delivery service to spoof recipients.

In this attack scenario, an attacker who successfully exploited the vulnerability could run arbitrary code in the context of the current user (recipient). If the user is logged on with the administrative user rights, the attacker could take control of the affected system. 
The attacker could then install programs; view, change, or delete data; or create new accounts with full user rights. 
As we can seen, users whose accounts are configured to have fewer user rights on the system could be less impacted than users who operate with administrative user rights.

\begin{figure*}[t]
    \centering
    \includegraphics[width=\linewidth]{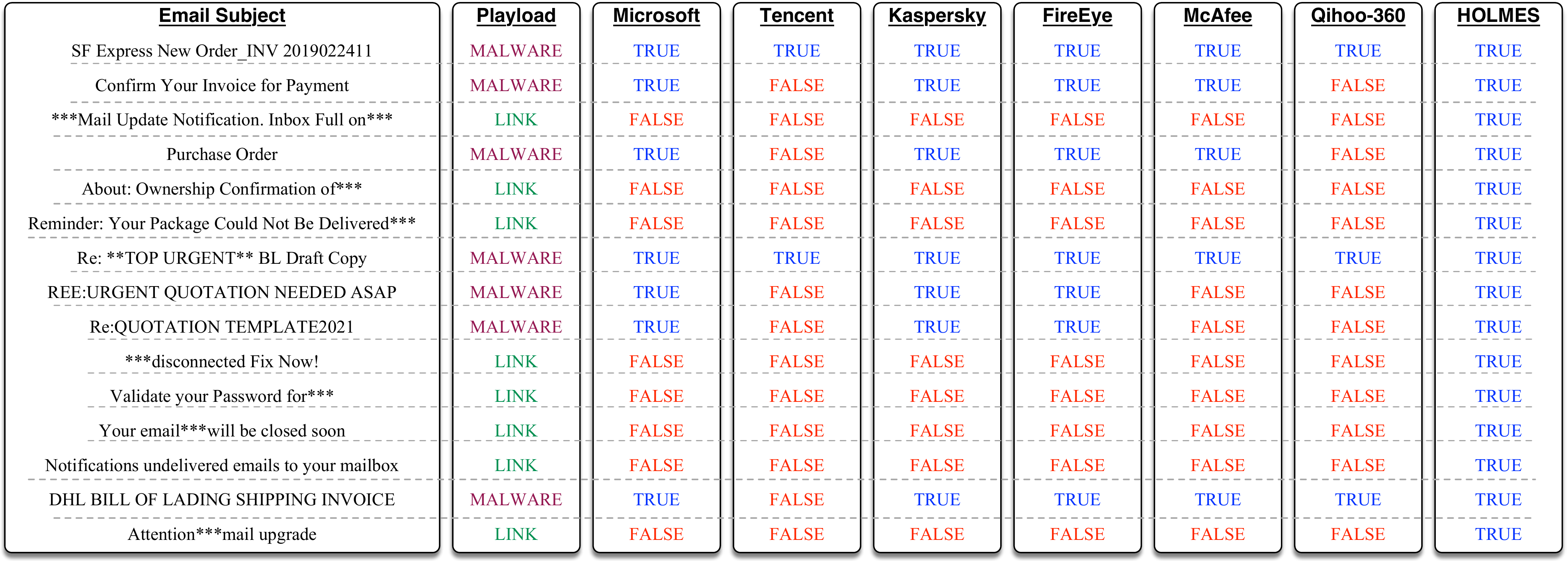}
    \caption{Detection Result of \textsc{Holmes} Compared with Other Detectors in VirusTotal Enterprise Version}
    \label{fig:comparions}
\end{figure*}

\subsubsection*{\textbf{Case B: Fake IT Security Alert}}\label{eml2}
Fig.~\ref{fig:mail_samples}(b) shows the execution flow of a malicious email that uses a deceptive subject named ``New Sign-in Attempt", aiming to spoof recipients to change their email account password.
Once the recipient clicks the button of ``Update security settings", the web page will be redirected to the phishing website:  \url{https://controladmin.7m.pl/login[.]html?#xxx@xxx.gov.xx}, which induces the victim user to type in the username and password.
The web page will, in the end, return to the enterprise homepage that the victim user works.

On the hacker side, the back-end server will receive the event log of the failed login attempts from the victim user, and then record the username and password.
Hence, the hacker can use the legitimate email account to sign in, such as web page or email server, and can even further send an elaborately crafted phishing email to a person who is the victim's frequent contact, 
which is hard to be detected by most security products.

\subsubsection*{\textbf{Case C: Phishing from a Legitimated Website}}
The malicious execution flow shown in Fig.~\ref{fig:mail_samples}(c) is similar to the attack 
shown in Fig~\ref{fig:mail_samples}(b) in that it also has a link embedded in the mail content for phishing campaign.
However, the phishing link \url{https://armonaoil.com/admin/images/npgtr/newsl/potcpanel[.]html} is from a legitimate website rather than from a personally created malicious website, which indicates that the legitimate website has been compromised for the use of darknet market\footnote{The darknet is most often used for illegal activities such as black markets, illegal file sharing, and the exchanging of illegal goods or services.}.

By further analysis, we found that the enterprise in this case indeed opened the cPanel web hosting server for public access, which was vulnerable to the brute-force attack and remote external control.
Furthermore, we examined the recent activities on some popular darknet markets and found that more than 3,000 sites of cPanel accesses were selling in the darknet market (raidforums) since 2020-11-22.
Hence, it can be confirmed that the email attack is a phishing campaign caused by the third-party information leakage.

\subsection{A Comparative Study}
To evaluate the detection capability of \textsc{Holmes}, we compare it with some commercial email detectors that are offered by six key security vendors in VirusTotal: Microsoft, Tencent, Kaspersky, FireEye, McAfee and Qihoo-360.
We select 15 malicious emails as testing samples. They either contain a phishing link or have a malware infected attachment.
All the testing samples were collected from the real-world threat hunting during the whole December month in 2020, and these samples had bypassed the detection of the enterprise anti-spam gateway and successfully detected by \textsc{Holmes}. More examples can be found in Table~\ref{tab:examples}.
The comparison is to demonstrate the proportion of highly-concealed malicious emails, which the other commercial tools still cannot to discover.
The result shows the evidence and reason why we still need a behavioural anomaly detection tool like \textsc{Holmes} for anomalous email detection.

The comparison table is given in Fig.~\ref{fig:comparions}, where the email subjects representing the 15 malicious emails are listed in the first column and the rest columns are the detection results from the commercial detectors and \textsc{Holmes}.
A FALSE value from a detector on a malicious email indicates that the detector failed to identify the malicious email. 

For the detectors of Microsoft, Kaspersky and FireEye,
we can see a good performance on detection of those malicious emails that 
contain malware infected attachments.
However, 
they fail to detect the malicious emails that contain phishing links. 
Based on the further analysis by our security experts, most of the phishing domains have been registered no more than three months and some of them are even from legitimate known enterprises.
Moreover, all the phishing links include a specific URL to access the particular crafted phishing web page under the domain name that is shortly expired in around three days. Such a short-lived situation
significantly increases the difficulty of anomaly detection.

From the comparison table, we can also see that
McAfee demonstrates a moderate detection rate on the malicious emails that contain malware infected attachments.
Similar to Microsft, Kasperly and FireEye, McAfee also cannot detect the malicious emails that contain a newly registered phishing link.

Compared to all above detectors, Tencent and Qihoo-360 have a low detection rate. Among the 15 malicious emails, only two are detected by Tencent and three by Qihoo 360.

We would clarify that, the detection engines used for the comparison are supplied by the VirusTotal Enterprise Service.
Since the version of the detectors may not be the same used in their commercial products, we would state that the comparison result cannot completely indicate the detection capability of their latest versions in the commercial products.



\begin{table*}
   \caption{\label{tab:examples}Samples of Phishing Emails Detected by \textsc{Holmes} on 2021-10-26.}
   \centering
\begin{tabular}{ |p{.26\linewidth}|p{.26\linewidth}|p{.26\linewidth}|  }

\hline
\multicolumn{3}{|c|}{Phishing Emails Detected by \textsc{Holmes} on 2021-10-26} \\
\hline
Subject: You have an outstanding payment. & Subject: Please confirm your email account. & Subject: Incoming mails has blocked! \\
\includegraphics[width=\linewidth]{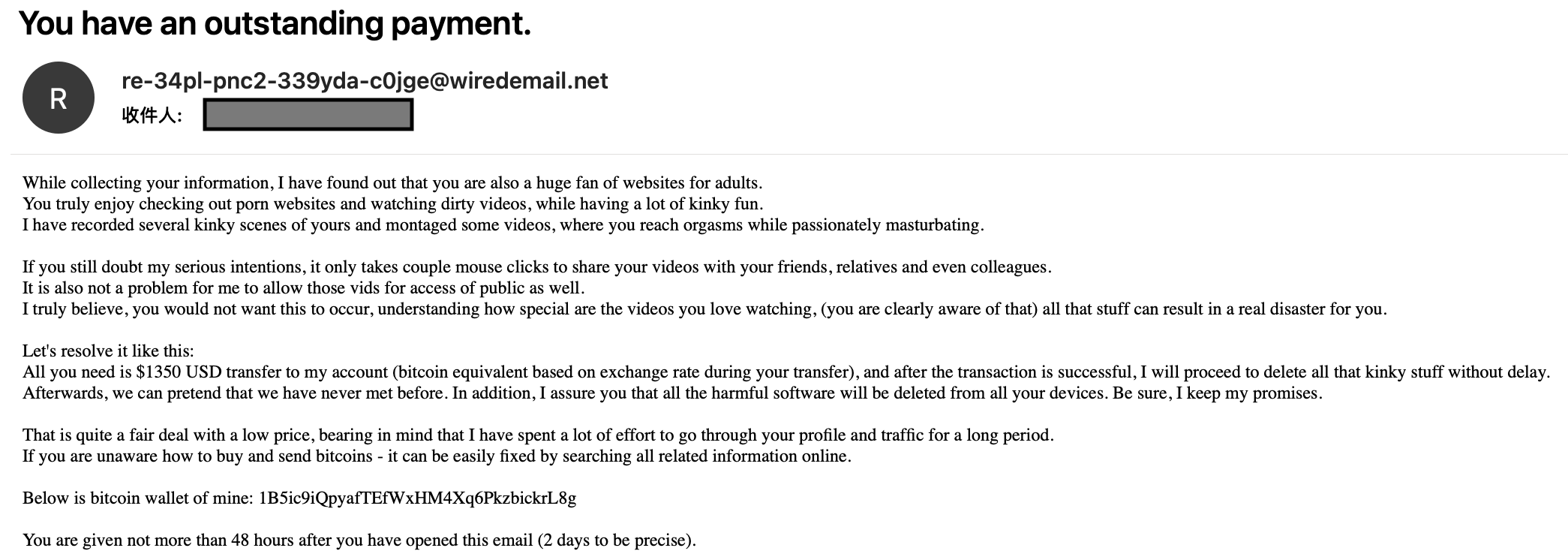} & \includegraphics[width=\linewidth]{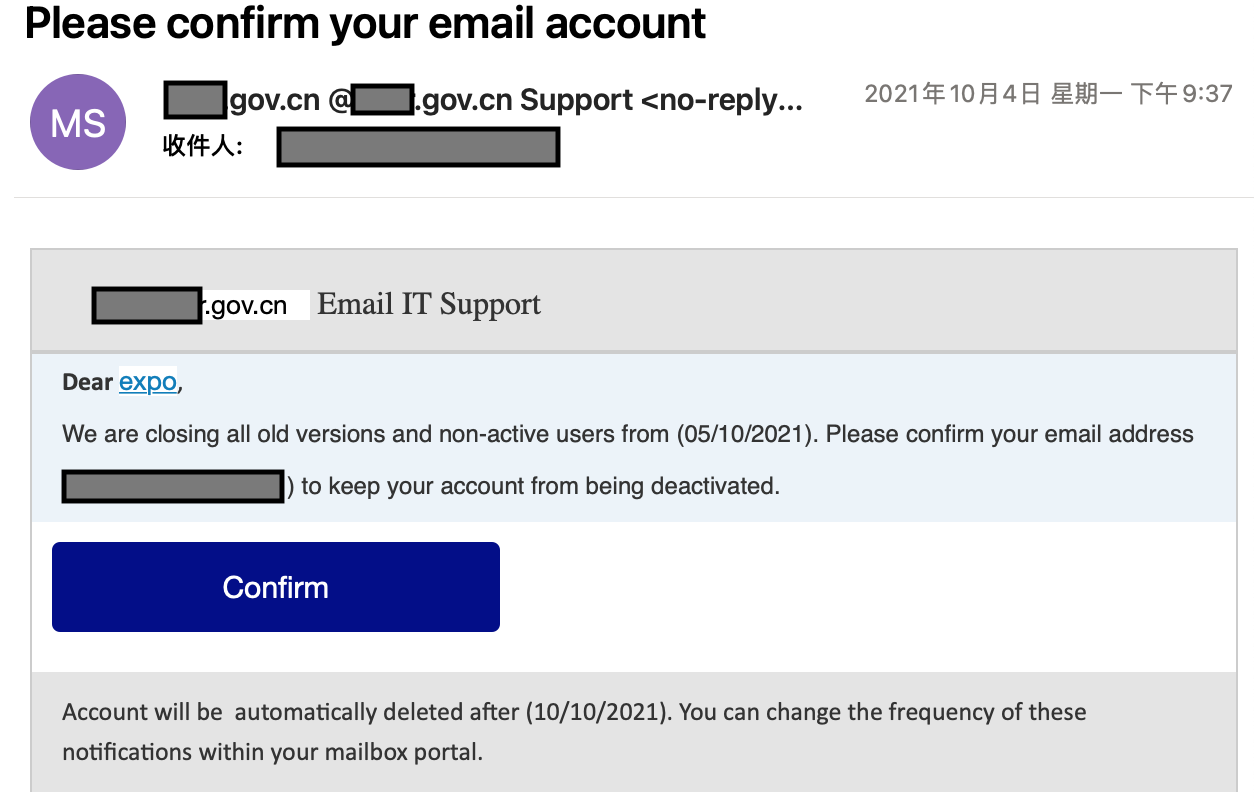}   & \includegraphics[width=\linewidth]{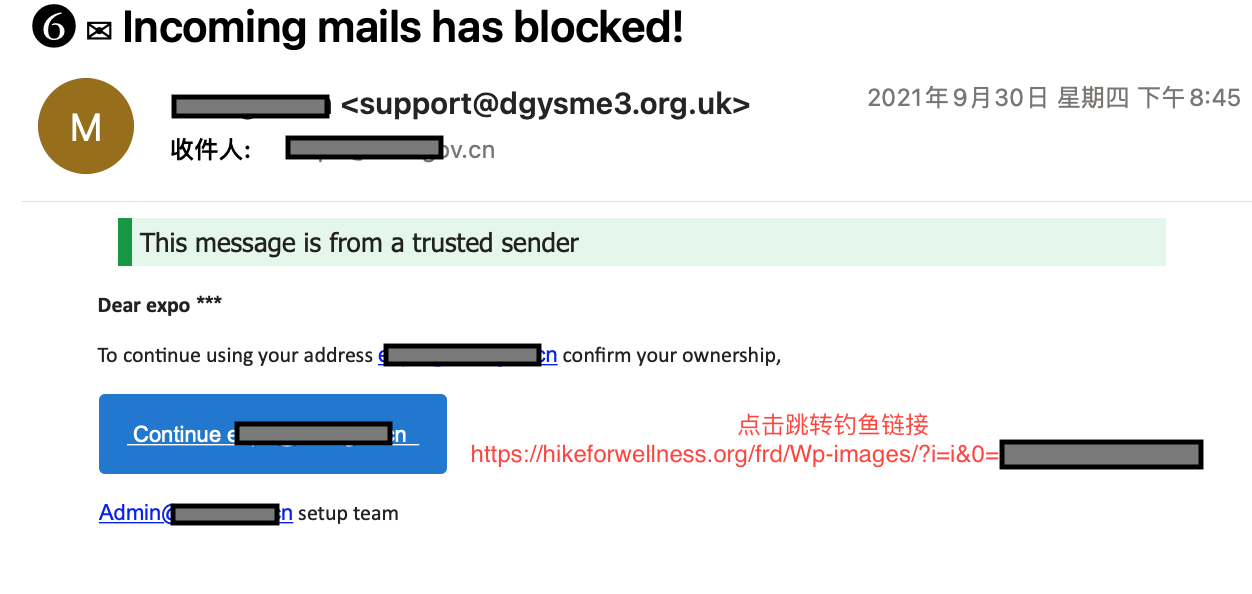} \\
\hline
Subject: Closure Of Mailbox Notice ! & Subject: Security notice - Immediate action required. & Subject: Security notice - Immediate action required. \\
\includegraphics[width=\linewidth]{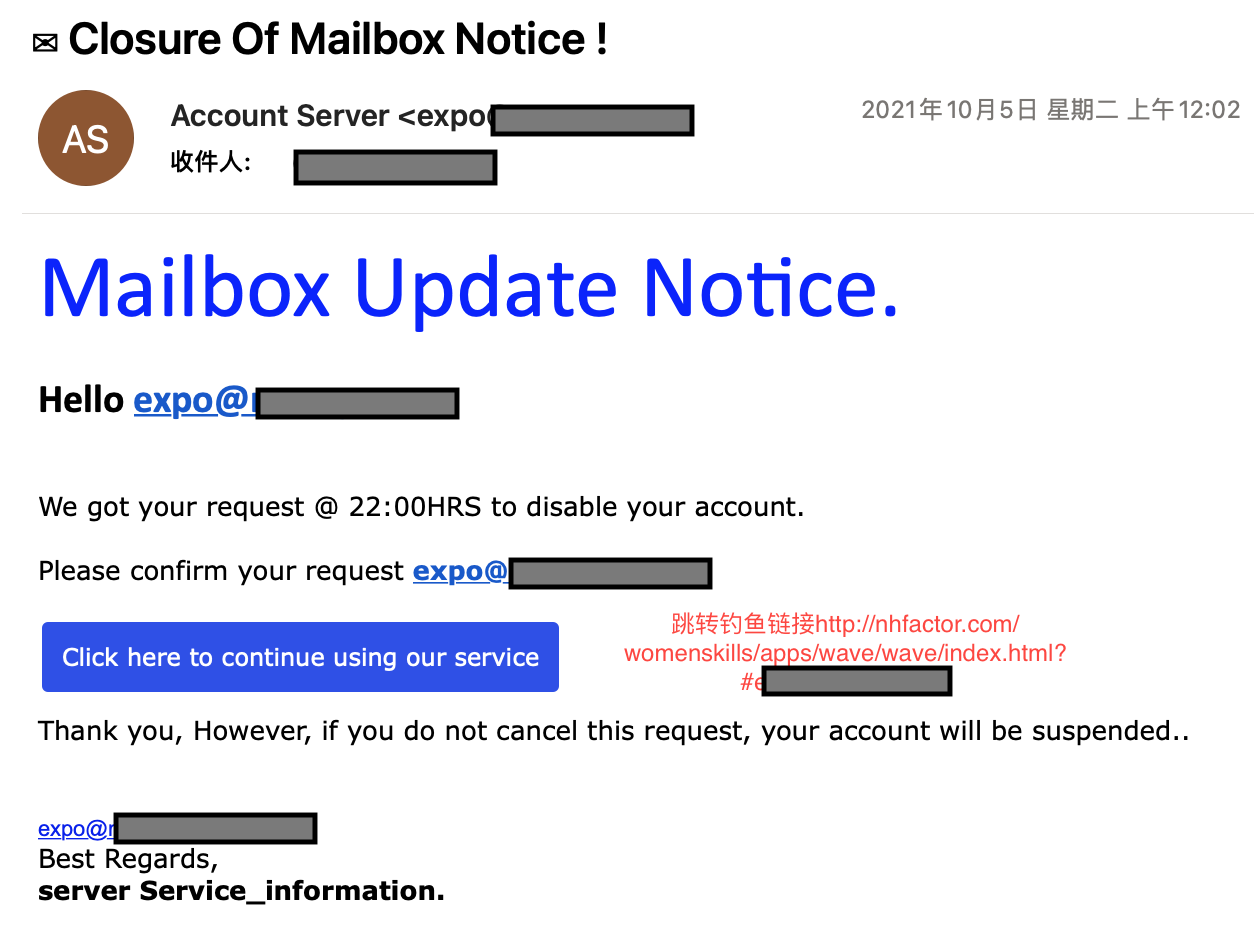} & \includegraphics[width=\linewidth]{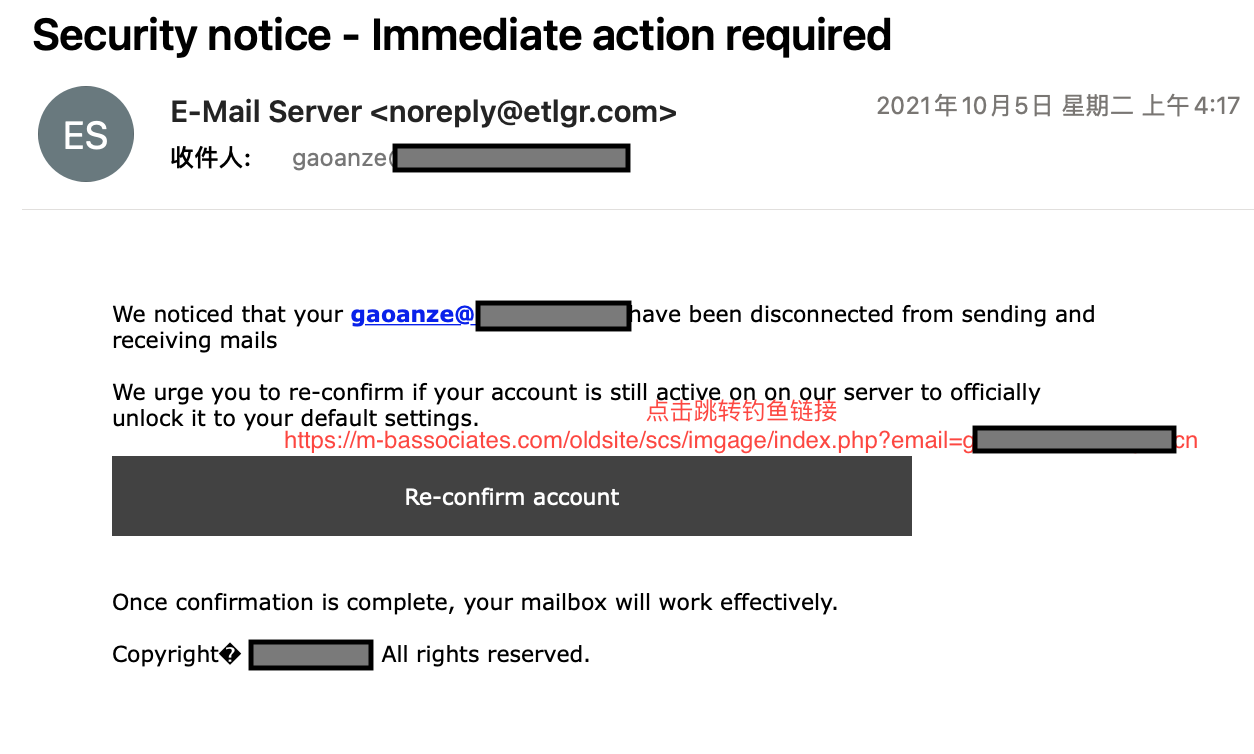}   & \includegraphics[width=\linewidth]{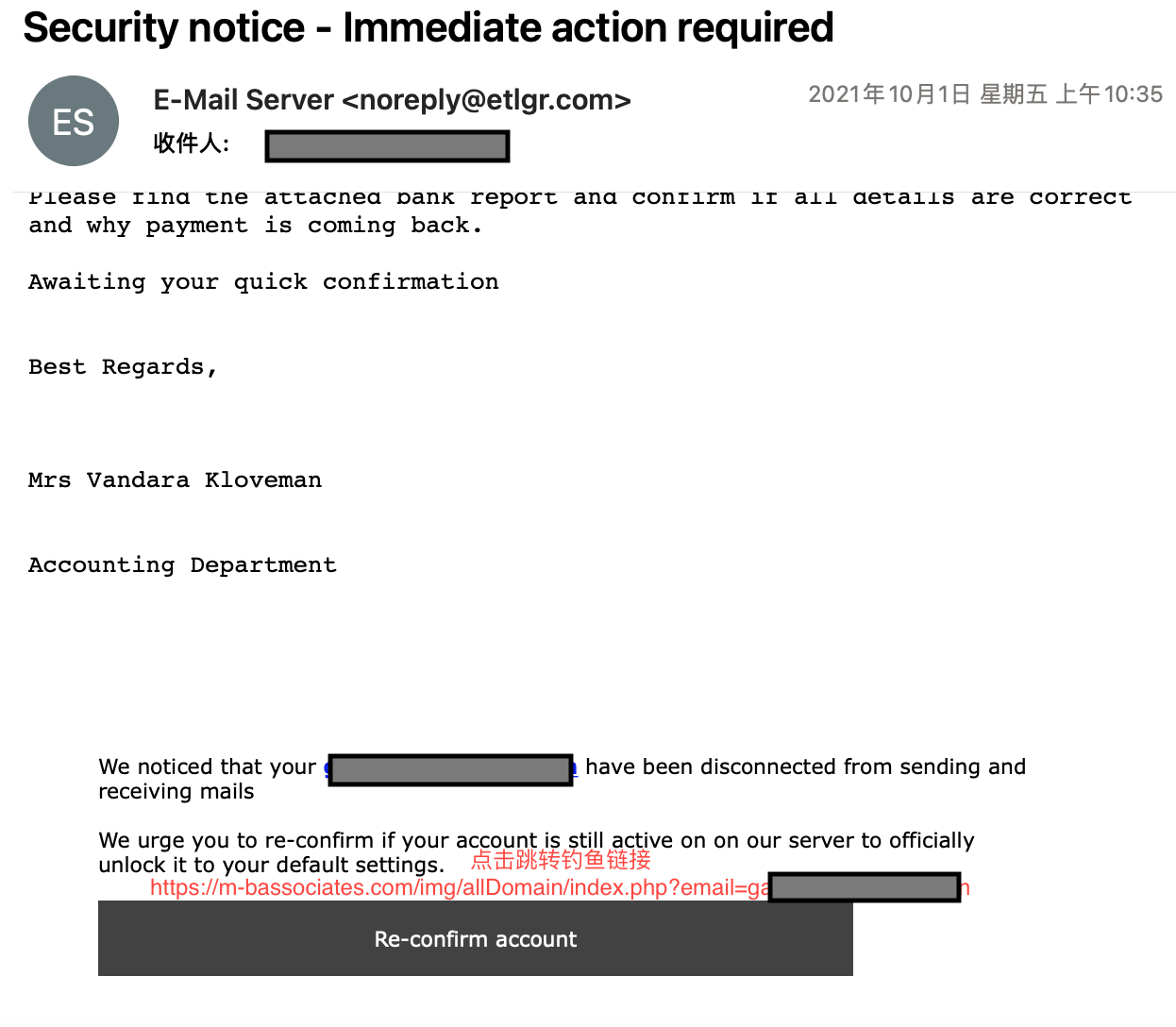} \\
\hline
Subject: xxx@xxx.gov.cn Protection & Subject: Email Account Security Alert Request. & Subject: Notification: xxx@xxx.gov.cn Disk is full. \\
\includegraphics[width=\linewidth]{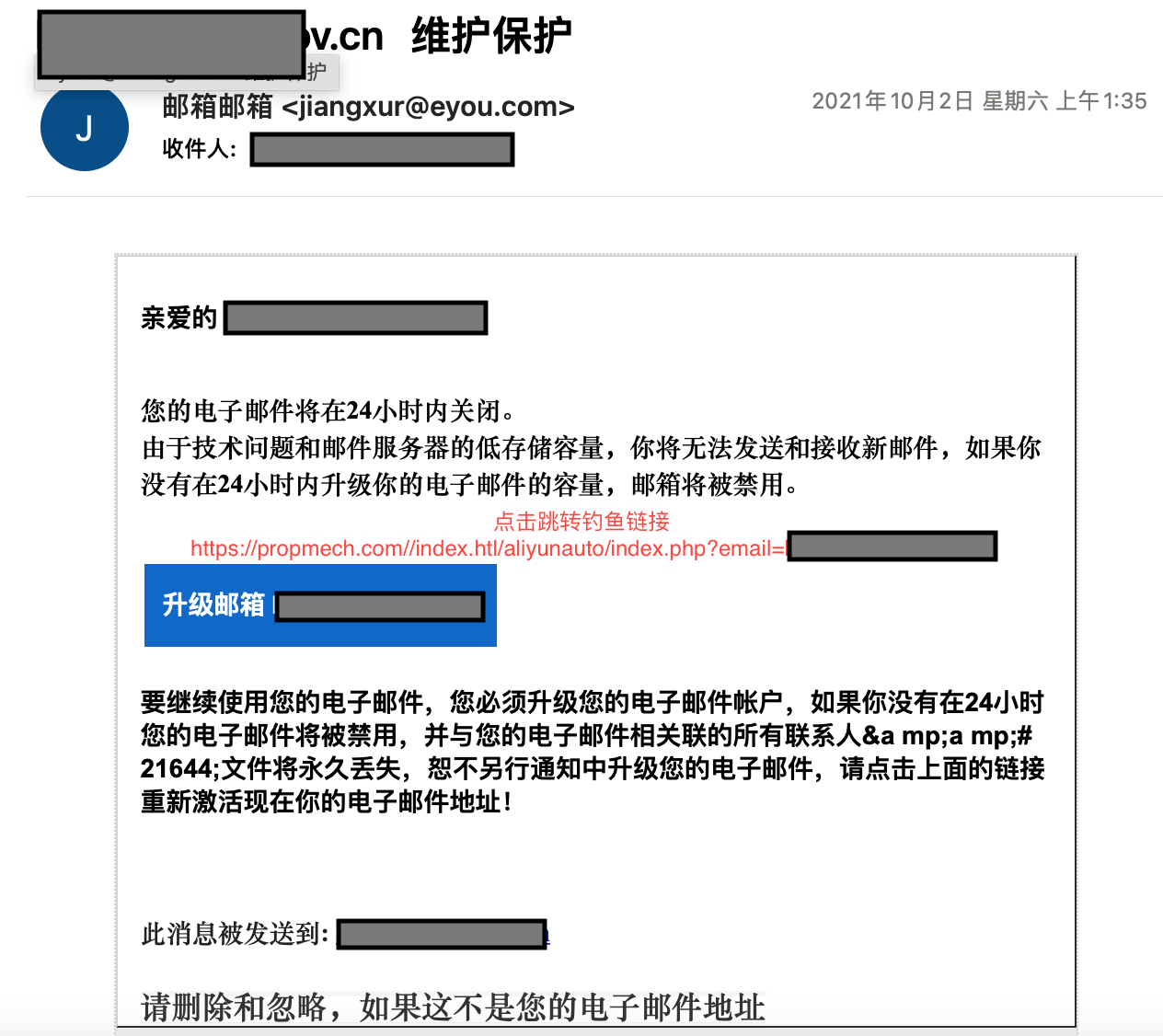} & \includegraphics[width=\linewidth]{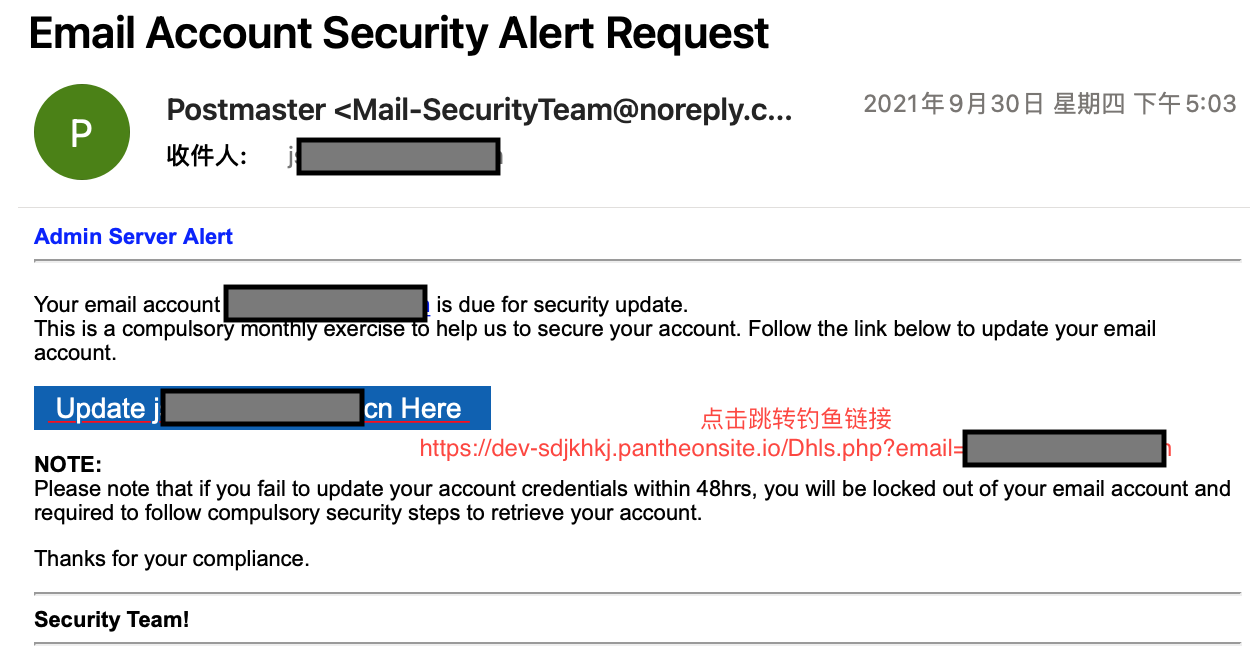}   & \includegraphics[width=\linewidth]{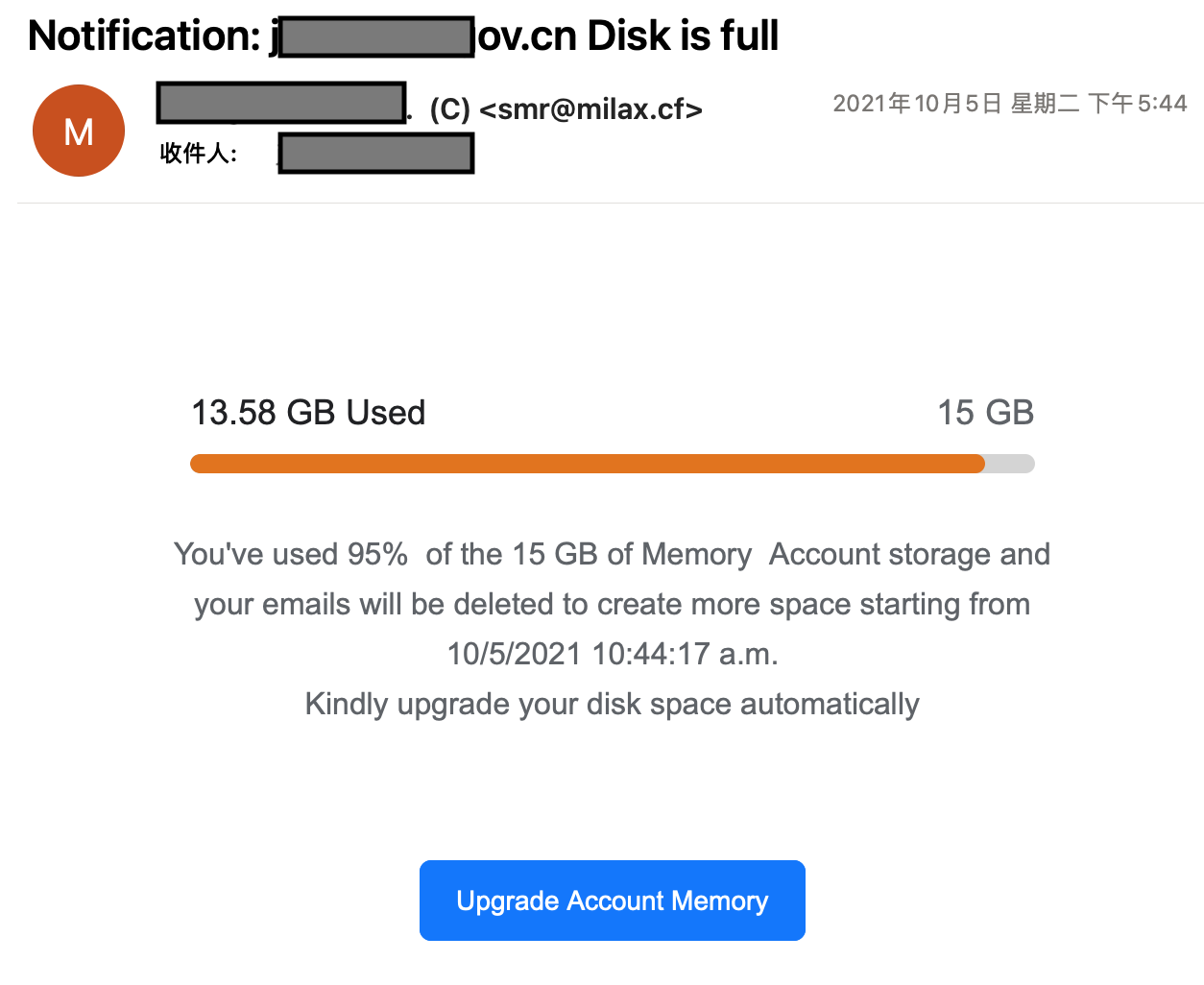} \\
\hline
Subject: [WARNING!!!]Your Email Account Is About To Be Terminated. & Subject: xxx@xxx.gov.cn Password Update. & Subject: One time verification. \\
\includegraphics[width=\linewidth]{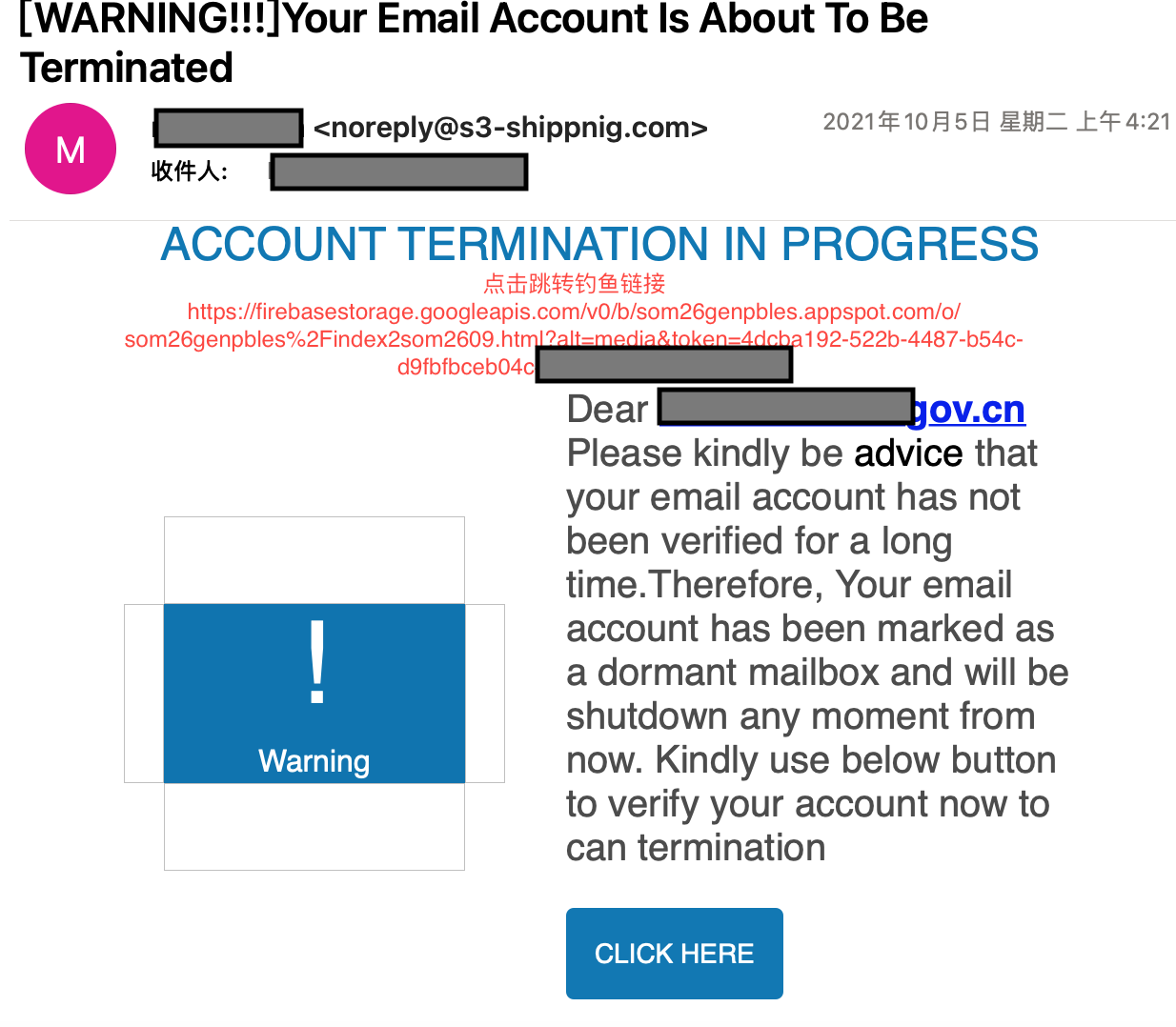} & \includegraphics[width=\linewidth]{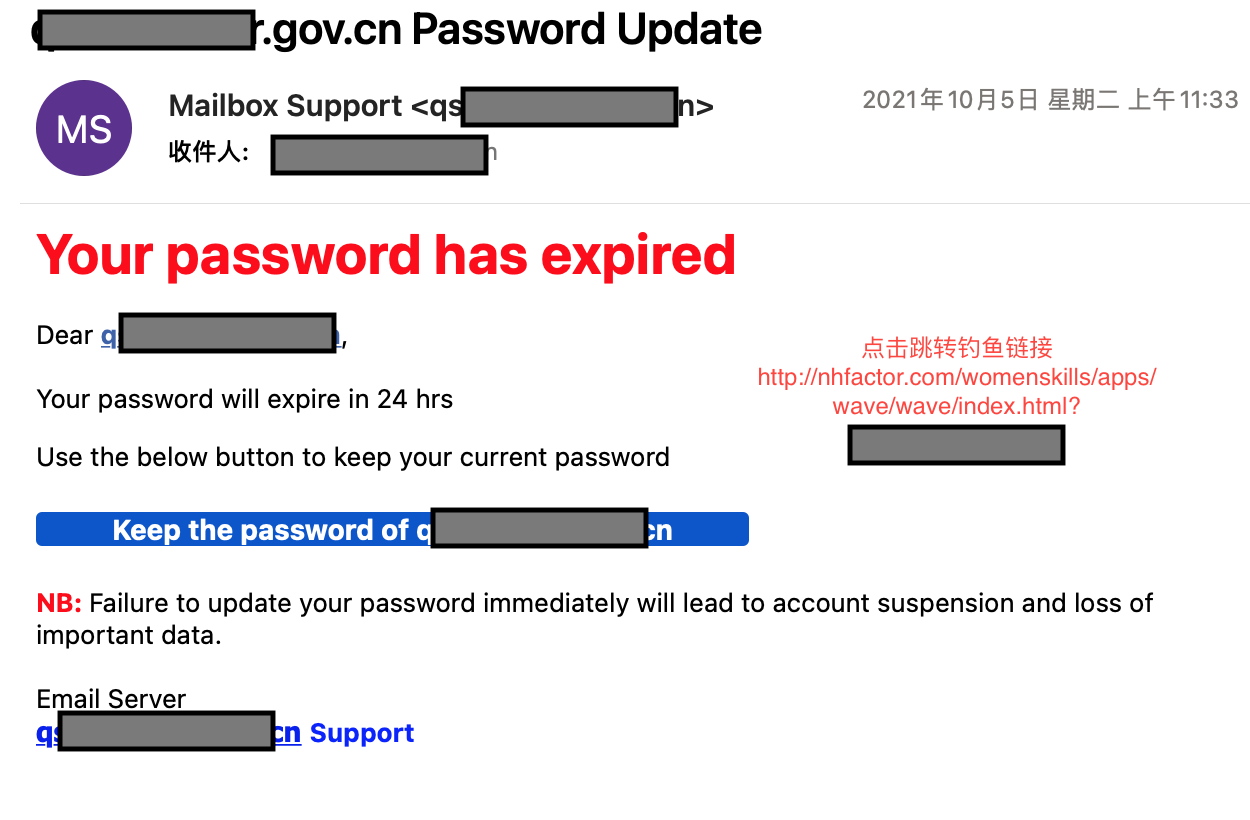}   & \includegraphics[width=\linewidth]{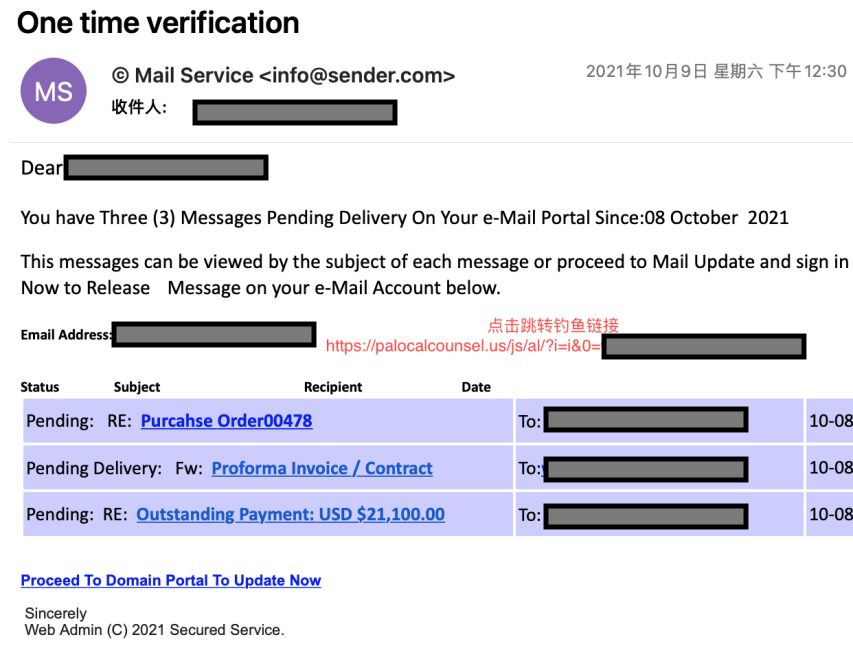} \\
\hline

\end{tabular}
\end{table*}

\section{Enhancements in the Latest Implementation}
\label{sec:implementation}

\begin{figure}[t]
    \centering
    \includegraphics[width=.97\linewidth]{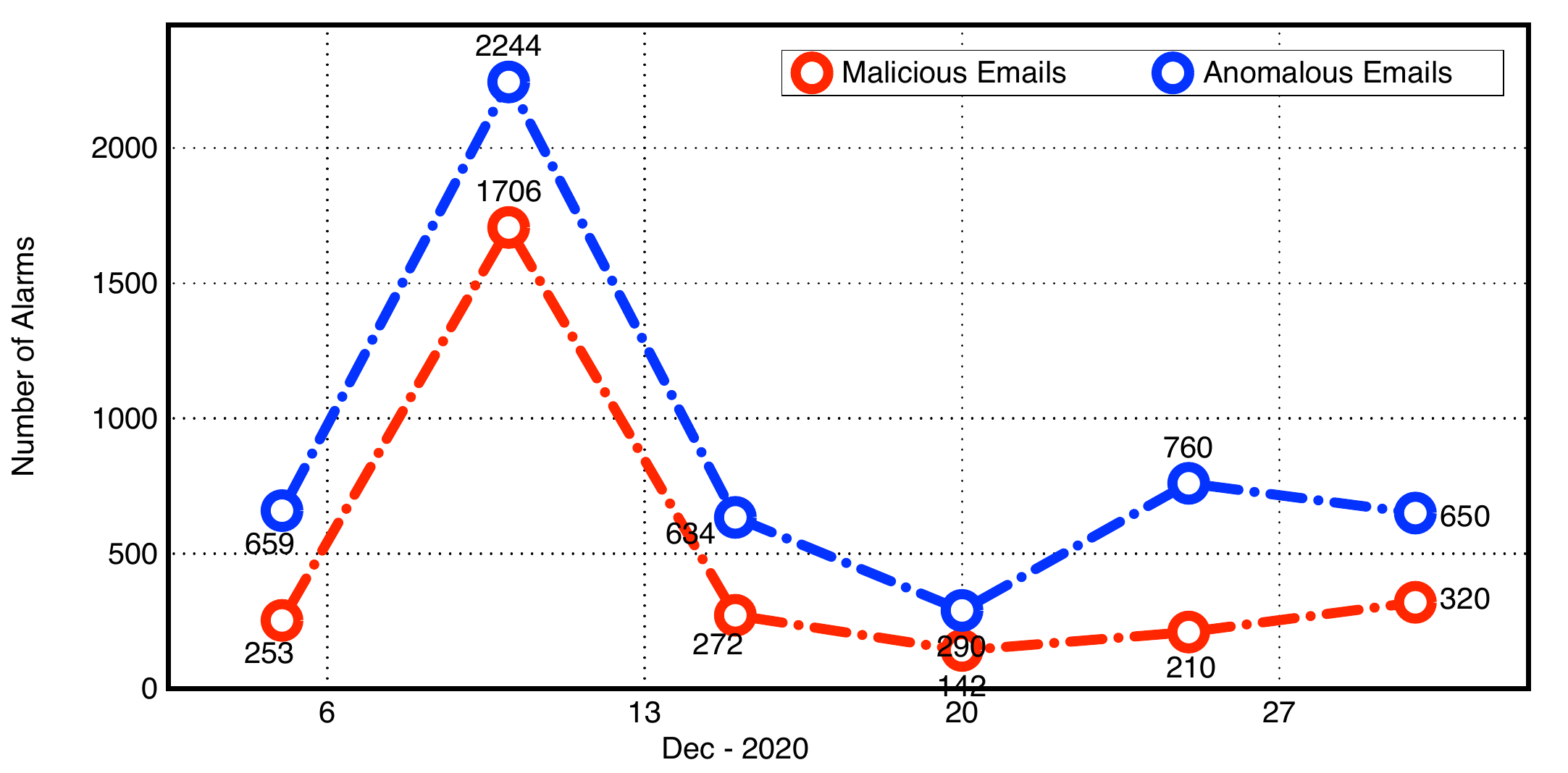}
    \caption{Detection Performance (December 2020)}
    \label{fig:dec}
\end{figure}

\begin{figure}[t]
    \centering
    \includegraphics[width=.97\linewidth]{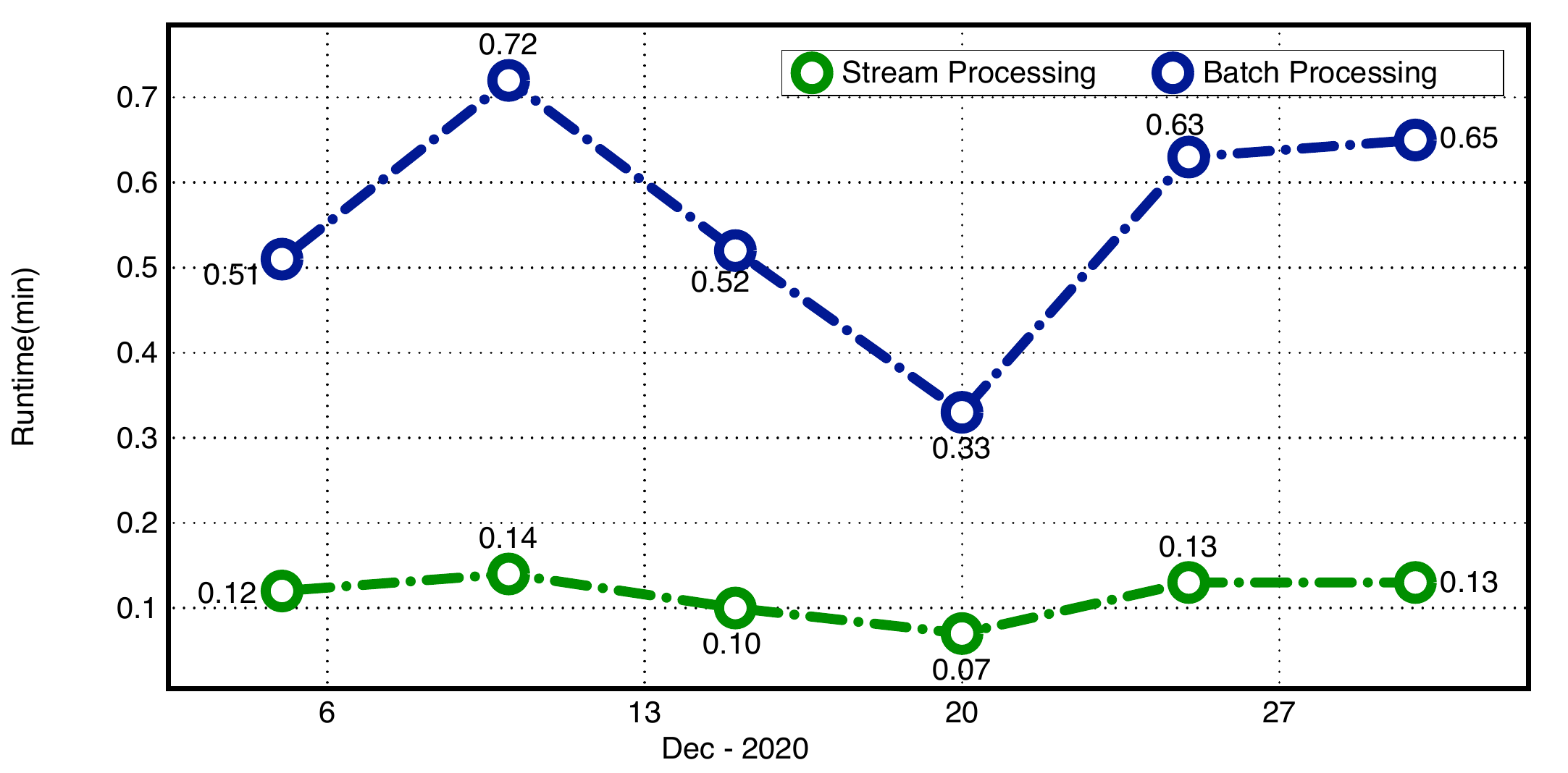}
    \caption{Stream Processing \& Batch Processing}
    \label{fig:stream}
\end{figure}

After the first deployment to the enterprise environment, as mentioned in the above section, \textsc{Holmes} has been upgraded with a few enhancements. 
In the latest version of \textsc{Holmes}, we rebuild the code warehouse that makes \textsc{Holmes} more efficient to discover anomalies in a much smaller rolling time window.
The improvement is achieved by moving the data query system from Elastic-Search (ES) server to the real-time Kafka computing platform.
The main difference between ES and Kafka is the way the data is processed. ES uses batch processing whereas Kafka uses stream processing, and the stream processing is more timely and efficient. 
The advantage of using Kafka is that \textsc{Holmes} can detect anomalies in less than one minute without the risk of server crash, significantly reducing the computing consumption.
Fig.~\ref{fig:dec} and Fig.~\ref{fig:stream} respectively shows the detection performance of \textsc{Holmes} in December 2020 and the run-time performance improvement after switching batch processing to stream processing. 
Furthermore, use of Kafka can also help our security analysts to better schedule time for threat identification and improve the efficiency of threat responses.

\section{Conclusion}
\label{sec:conclusion}

In this paper, we introduce \textsc{Holmes}, a lightweight semantic based anomalous email detector, which can effectively discover malicious emails in the real-world cyber threat hunting.
\textsc{Holmes} also demonstrates a viable solution that successfully transfers AI technology to the cyber security field and makes an excellent trade-off between the cost of algorithmic consumption and the detection performance. 

We measure the performance of \textsc{Holmes}, and compare its detection capability with several well-known commercial detectors offered by the security companies in VirusTotal.
Our evaluation result shows that, on the use of cyber threat hunting, \textsc{Holmes} significantly outperforms those commercial products in a range of malicious attack scenarios, which demonstrates its practical values in the commercial competition.
Our current evaluation is based on one-month data. As future work, we will extend our investigation on a large scale dataset that is collected over a long test period and covers emails of different languages.

\renewcommand{\baselinestretch}{1.17}
{\footnotesize \bibliographystyle{ieeetr}
\bibliography{sample}}

\end{document}